\documentclass[aps,prl,superscriptaddress,twocolumn,citeautoscript,showpacs,floatfix]{revtex4}
\usepackage{amssymb} \usepackage{amsfonts} \usepackage{amsmath}
\usepackage{graphicx} \usepackage{ulem}\usepackage{latexsym}

\newcommand{\comment}[1]{}
\def \be{\begin{equation}}
\def\ee {\end{equation}}
\def \ber{\begin{eqnarray}}
\def\eer {\end{eqnarray}}

\def\bef{\begin{framed}}
\def\eef{\end{framed}\noindent}

\def \Av{{\bf A}}

\def \Ev{{\bf E}}

\def \rv{{\bf r}}

\def \zv{{\bf z}}
\def \pv{{\bf p}}

\def \kv{{\bf k}}
\def\pv {{\bf p}}

\begin{document}
\title{Inverse Edelstein Effect}
\author{Ka Shen}
\affiliation{Department of Physics, University of Missouri,
  Columbia MO 65211}
\author{G. Vignale}
\affiliation{Department of Physics, University of Missouri,
  Columbia MO 65211}
 \affiliation{Italian Institute of Technology at Sapienza and Dipartimento di Fisica, Universit\`a ``La SapienzaÓ, Piazzale Aldo Moro 5, 00185 Rome, Italy}
 \author{R. Raimondi}
\affiliation{CNISM and Dipartimento di Matematica e Fisica, Universit\`a Roma Tre,
Via della Vasca Navale 84, 00146 Rome, Italy}
\begin{abstract}
We provide a precise microscopic definition of the recently observed  ``Inverse Edelstein Effect" (IEE), in which  a  non-equilibrium spin accumulation in the plane of a two-dimensional (interfacial) electron gas drives an electric current perpendicular to its own direction.  The drift-diffusion equations that govern the effect are presented and applied to the interpretation of the experiments.
\end{abstract}
\date{\today} \maketitle
{\it Introduction} - 
The spin Hall effect (SHE) and the inverse spin Hall effect (ISHE) are  well established phenomena~\cite{Dyakonov71,Hirsch99,Zhang00,Murakami03,Sinova04,kato04,Murakami05,Wunderlich05,Engel07,Hankiewicz09,Jungwirth12}, which play an important role in experimental spintronic devices~\cite{Awschalom02,Zutic04,Fabian07,Awschalom07,Wu10,Tsymbal11}.  In the SHE an electric current $J_x$, driven by an electric field $E_x$ produces a $z$-spin-current in the $y$ direction, denoted by $J^z_y$.  In the ISHE, which is the Onsager reciprocal of the SHE, a spin current $J^z_y$, driven by a ``spin electric field" $E^z_y$  produces an electric current $J_x$ in the $x$-direction.   
Both effects are  characterized by the  spin-Hall conductivity, which can be as large as $10^5$~($\Omega$ . m)$^{-1}$ in bulk metals like Pt~\cite{Seki08,Liu12}.  The SHE plays an important role in technology as a source of spin currents  that can, for example,  excite a spin wave in a ferromagnet or flip the magnetization of an element of a spin valve structure~\cite{Liu12}.  Likewise, the ISHE has been exploited for the detection of spin currents~\cite{Valen06,Seki08,Uchida08}.

Another well-known effect, intimately related to the SHE, is the so-called Edelstein effect (EE)~\cite{Edelstein90}  (notice, however, the paper published at about the same time by Lyanda-Geller and Aronov~\cite{Aronov89}).  In this effect, a steady current $J_x$, driven by an electric field $E_x$ produces a steady non-equilibrium spin polarization $S^y$.  The effect has been observed experimentally~\cite{Kato04,Silov04} and can be understood, on a basic level, as the result of the effective magnetic field (due to spin-orbit coupling) ``seen" by the drifting electrons in their own reference frame.  


After much theoretical work in the past decade, an intuitive and useful drift-diffusion theory of the SHE,  ISHE, and  EE has recently emerged (see Refs.~[\onlinecite{Gorini10,Raimondi12, Gorini12}]).  This theory is firmly grounded in  quantum kinetic equations and diagrammatic calculations for systems in which the spin-orbit interaction is linear in $\kv$ and can therefore be described by an SU(2) vector potential.  In the meanwhile, no attention has been paid to the ``inverse Edelstein effect" (IEE), by which we mean the Onsager reciprocal of the normal Edelstein effect.  
Although the spin-galvanic effect observed almost a decade ago by Ganichev {\it et al.}~\cite{Ganichev02} 
 in GaAs may be interpreted as  a 
 manifestation
of the IEE,
to the best of our knowledge the latter 
has  been first introduced and  quantitatively characterized
in a very recent experimental paper by Rojas S\'anchez {\it et al.}~\cite{Sanchez13}. 
In this paper we provide a precise theoretical characterization of this effect and introduce the drift-diffusion equations that describe it. 

At first sight, the IEE is puzzling:  a static magnetic field $B^y$, which couples linearly to the spin density $S^y$, will  not create an electric current in the $x$-direction: rather, it will change the value of the equilibrium spin polarization $S^y$.
This reasoning fails to recognize the essential difference that exists between the spin polarization created by a static magnetic field in equilibrium  and the non-equilibrium spin polarization that arises from a steady spin injection.  In both cases the spin polarization is constant in time, but it is only the non-equilibrium one that drives an electric current (IEE). 
The theoretical problem is  to identify the mechanical field that is reciprocal, in the sense of Onsager's reciprocity relations, to the electric field $E_x$.  This is the field that describes the physical process of spin injection -- an effect that is more commonly treated as a source term in the Boltzmann collision integral.

It will be shown below that the sought ``spin injection field" is simply a magnetic field that varies linearly in time.  Such a field produces  a spin density that, at each instant of time, lags slightly behind the instantaneous equilibrium spin density, by an amount proportional to the spin relaxation rate.   The difference between the true spin polarization and the instantaneous equilibrium polarization  is the proper non-equilibrium spin density injected by the field. 
The efficacy of the injection is thus limited by the spin relaxation time $\tau_s$~\cite{footnote1}.

\begin{figure}
\begin{center}
\includegraphics[width=0.9\linewidth]{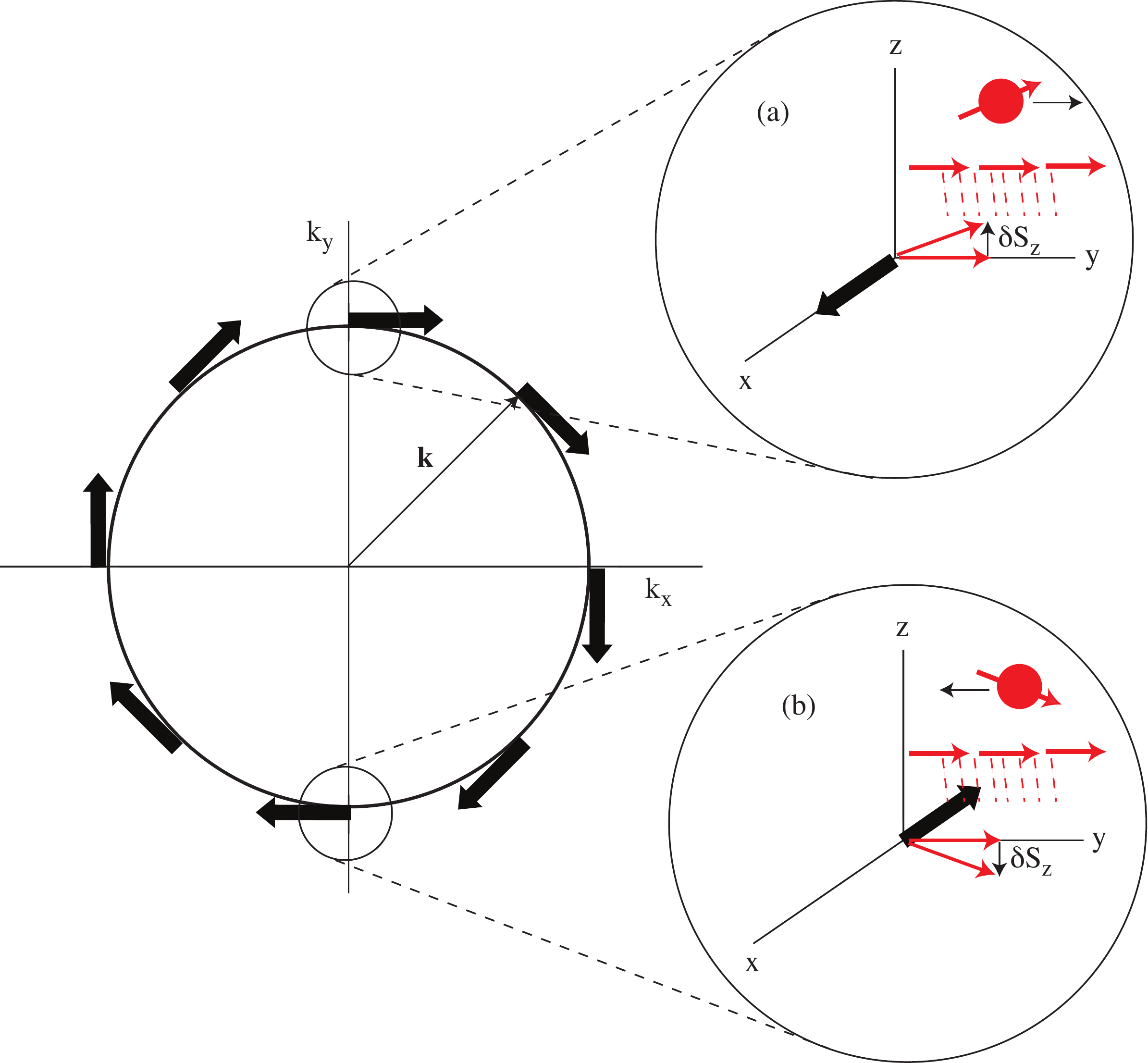}
\end{center}
\caption{
Schematic description of the Rashba model and of the generation of a spin current from injected spin in the y direction.  The thick arrows denote the Rashba field around which the spin (thin arrow) precesses.  The two insets (a) and (b)  zoom in on the dynamics of the spin near  points  $(0,k_F)$ and $(0,-k_F)$ in momentum space.  In the first case the spin, initially pointing in the $y$ direction, acquires a positive $z$-component and travels in the $+y$ direction.  In the second case, the spin acquires a negative $z$ component and travels in the $-y$ direction.}
\label{Fig1}
\end{figure}

Figure~\ref{Fig1} shows the qualitative picture of the IEE for the paradigmatic case of the Rashba spin-orbit coupling, i.e., a spin-orbit coupling of the form $\alpha (\zv \times \pv)\cdot {\boldsymbol \sigma}$, where $\zv$ is the unit vector perpendicular to the plane of the electrons,  $\alpha$ is the Rashba velocity, and $\pv$ the momentum of the electron.  
 If spin injection could be selectively done at fixed momentum,
 it would be, then,  obvious in order to produce a charge current along the $x$ direction to inject and extract electrons at $(-k_F,0)$
 and $(k_F,0)$, respectively, where the Rashba field  
  $\alpha (\zv \times \pv)$  aligns the spin in the $y$ direction (see Figure ~\ref{Fig1}). 
   This may happen at the surface of a topological insulator, where momentum and spin are locked
 \cite{Schwab2011}. However, in the Rashba model, $y$-polarized spin injection may occur at any momentum,
   irrespective of the direction of the internal field $\alpha (\zv \times \pv)$.
  Take for instance the apparently least favourable 
   case  at $(0,k_F)$ or $(0,-k_F)$.
Under the action of the  Rashba field $\alpha (\zv \times \pv)$ the spin  of an injected electron, initially pointing along $y$, acquires a $z$-component that is positive or negative according to whether  $p_y$ is positive or negative. The resulting correlation between $p_y$ and $S^z$ is the signature of a spin current $J^z_y$.  At this point the regular inverse spin Hall effect takes hold~\cite{Dyakonov07}, converting part of the spin current to a perpendicular charge current $J_x$.
  This shows, as we will derive later in a quantitative way, that spin density, in the Rashba model,  is intimately related to both spin and charge currents in such a way that
the final result  is a direct proportionality between $J_x$ and the incoming current of $S^y$ spin:
\be\label{Equation1}
J_x=\lambda_{IEE} J^y_s,
\ee
where $\lambda_{IEE}$ connects an areal current density to a volume spin current density, and therefore has the dimensions of a length if, as we do in this paper, we express both charge and spin in the same units (this is achieved by multiplying the spin by $-\hbar/e$).  In the simplest case of the pure Rashba model we find $\lambda_{IEE}=\alpha \tau$, where $\tau$ is the momentum relaxation time.

In this Letter, we  provide a precise formal definition of the IEE in terms of the Kubo formula for the response of the current $J_x$  to the $\dot B_y$ field, which is the Onsager reciprocal of the electric field $E_x$.  We also develop the drift-diffusion theory of the IEE  along the lines of Refs.~[\onlinecite{Gorini10,Raimondi12, Gorini12}] in the presence of both intrinsic and extrinsic (but linear in $\kv$) spin orbit coupling.  In the pure Rashba limit these equations yield Eq.~(\ref{Equation1}).  Lastly,  we make contact with the recent experimental work on the generation of a charge current by  spin injection into a Ag-Bi interface~\cite{Sanchez13}, and 
 identify
the proper relaxation time to be used in the expression for $\lambda_{IEE}$ 
as the momentum relaxation time.
 
{\it Formal definition of IEE} --  
The direct Edelstein effect is defined by the proportionality
\be
S^y (\omega) = \sigma_{DEE} (\omega) E_x (\omega)\,,
\ee
where we have allowed a periodic variation of the field and the induced density at a frequency $\omega$. 
To formalize the calculation of the ``Edelstein" conductivity $\sigma_{DEE} (\omega)$ we introduce the Kubo response function of the homogeneous spin density $S^y$ to a vector potential $A_x$, such that $E_x=-\dot A_x$.   Since $A_x$ couples linearly to the current density we denote this response by
$
\langle\langle \hat S^y;\hat J_x\rangle\rangle_\omega,
$ 
where $\hat J_x$ is the operator of the physical current (obtained by differentiating the Hamiltonian with respect to $A_x$) and $\hat S^y$ is the operator of the spin density.  The double bracket denotes the  Kubo product $\langle\langle \hat A;\hat B\rangle\rangle_\omega\equiv -\frac{i}{\hbar}\int_0^t \langle[\hat A(t),\hat B(0)]e^{i\omega t}dt$.  Since the electric field is related to the vector potential by $\Ev(\omega)=i\omega\Av(\omega)$ we immediately see that
\be
\sigma_{DEE} (\omega)=\frac{\langle\langle \hat S^y;\hat J_x\rangle\rangle_\omega}{i\omega}\,.
\ee
In the d.c. limit the numerator vanishes by gauge invariance because a static and uniform vector potential does not change $S^y$.  Then we obtain the d.c. Edelstein conductivity
\be
\sigma_{DEE} (0)=\lim_{\omega\to0} \frac{\Im m \langle\langle \hat S^y;\hat J_x\rangle\rangle_\omega}{\omega}\,,
\ee
which is a real quantity.

The inverse Edelstein effect is similarly defined by the proportionality
\be
J_x(\omega)=\sigma_{IEE}(\omega)[g\mu_B \dot B_y(\omega)]\,,
\ee 
 where, as described in the introduction, $\dot B^y$ is the field that injects a non-equilibrium spin density $S^y$.
We then introduce the Kubo response function of the homogeneous current density to a magnetic field that couples linearly to the spin density, namely
$
\langle\langle \hat J^x;\hat S^y\rangle\rangle_\omega.
$ 
Noting that $-\dot  B^y(\omega)=i\omega B^y(\omega)$ yields
\be
\sigma_{IEE} (\omega)=\frac{\langle\langle \hat J_x;\hat S^y\rangle\rangle_\omega}{i\omega}\,.
\ee
 In the d.c. limit
 \be
\sigma_{IEE} (0)=\lim_{\omega\to0} \frac{\Im m \langle\langle \hat J_x;\hat S^y\rangle\rangle_\omega}{\omega}\,.
\ee
The Onsager relation, namely the equality of the conductivities $\sigma_{DEE}$ and $\sigma_{IEE}$ follows immediately from well-known properties of the Kubo product, for a system governed by a time-reversal invariant hamiltonian.
This establishes the IEE as described above as the proper Onsager reciprocal of the standard Edelstein effect.

{\it Drift-diffusion theory} --  An elegant description of the direct spin Hall effect and Edelstein effect has recently been derived~ based on the methods of the quasi-classical Keldysh Green function technique~\cite{Gorini10,Raimondi12, Gorini12}.  We recall here the main aspects of  this description for the case of the two-dimensional electron gas with Hamiltonian
\be
\label{RH}
H=\frac{p^2}{2m}+V(\rv)+\alpha(p_y\sigma^x-p_x\sigma^y) +\lambda [\pv \times \nabla V(\rv)]  \cdot{\boldsymbol \sigma}
\ee
where $V(\rv)$ is the impurity potential.  The orbital motion takes place in the $(x,y)$ plane, while the spin is three-dimensional.  The last two terms represent the intrinsic and the extrinsic spin-orbit coupling respectively.  $\lambda$ is the extrinsic coupling constant, directly related to the square of the effective ``Compton wavelength" for the conduction band in which the 2DEG resides; $\alpha$ is the Rashba velocity, proportional to the electric field $E_z$ perpendicular to the plane, and approximately given by $\lambda E_z$.  
Since the complete formal derivation of the drift-diffusion theory presented below has been provided in Ref.~[\onlinecite{Gorini10}], we limit ourselves here to providing an heuristic justification.  We exploit the fact that the 
spin-orbit coupling  is linear in electron momentum $\pv$ and can therefore be represented by a constant SU(2) vector potential
$p^2/2m+\alpha (p_y\sigma^x-p_x\sigma^y)=(\pv +e {\bf A}^a\sigma^a /2)^2/2m$,  where the only non zero components are
 $-e A_y^x=e A_x^y=-2m\alpha$, 
The non Abelian character of the $SU(2)$ group entails the appearance of covariant derivatives
$ \nabla_i$  defined as follows:
\be\label{CovariantDerivative}
( \nabla_i O)^a =  \partial_i O^a-e\sum_{b,c}\epsilon_{abc}A^b_iO^c\,,
\ee
where $O^a$ is a generic vector function (in spin space) on which the derivative acts.
Notice that the second term in Eq.~(\ref{CovariantDerivative})  differs from zero even for a homogeneous function. An immediate consequence is the appearance
of a $SU(2)$ magnetic field different from zero even for uniform and constant vector potential.  It is given by the covariant curl of the SU(2) vector potential: $-e {\cal B}^a_i=({\rm i}/2)\epsilon_{ijk} \left[ A_j,A_k\right]^a$ 
with only non zero component $-e {\cal B}_z^z=-(2m\alpha)^2$. 
Once one accepts the above $SU(2)$ language, it is not too surprising that the coupled equations for charge 
and spin currents take the form
first introduced in Ref.~[\onlinecite{Gorini10}], namely
\begin{eqnarray}
\dot S^a&=&-\nabla_i J^a_i -\frac{\delta S^a}{\tau_{EY}},\label{SpinContinuity}\\
J_i^a&=&-D(\nabla_i \delta S)^a+{\sigma_s}E^a_i+\frac{e\tau}{2m}\epsilon_{ijk}J_j{\cal B}^a_k,\label{SU2SpinCurrent}\\
J_i&=&-D \partial_i n +\sigma_D E_i +\frac{e\tau}{2m}\epsilon_{ijk}J^a_j{\cal B}^a_k\label{SU2ChargeCurrent},
\end{eqnarray}
where $D=\frac{v_F^2\tau}{2}$ is the  diffusion constant,  $v_F$ is the Fermi velocity of the 2DEG,  $\sigma_D=\sigma_s=\frac{ne\tau}{m}$ 
is the Drude conductivity. 

 The first equation of the set is the continuity equation for the spin density.  The non-conservation of the spin, due to the action of the Rashba field, is taken into account via the replacement of the ordinary derivative by the $SU(2)$ covariant derivative,  whereas the additional spin relaxation due to impurity scattering is taken into account via the phenomenological Elliot-Yafet relaxation time $\tau_{EY}$~\cite{footnote_EY}.
 The quantity $\delta S^a=S^a- S^a_e $ is the deviation of the spin density from the instantaneous equilibrium spin density $S^a_e = \chi_{s}B^a(t)$,  where $\chi_{s}=-g\mu_B \frac{m}{ \pi}$ 
  is the static spin susceptibility. 

 The last two equations express the spin current and the charge current densities as sums of
{\it diffusive}, {\it drift} and {\it Hall}-like terms~\cite{footnote_ddh}.  Whereas in uniform circumstances the charge current does not have
 a diffusion contribution,  the spin current, due to the $SU(2)$ covariant derivative, does have an anomalous diffusion contribution -- the first term on the right hand side of Eq.~(\ref{SU2SpinCurrent}).  This anomalous diffusion  describes the spin current that arises from the precessional motion of the spins in the Rashba field.  Its origin is described qualitatively in Fig. 1.    

 Whereas the second term on the right hand side  of Eq.~(\ref{SU2SpinCurrent}) is the well known drift term due to the spin-electric field arising from the spin accumulation potential of classical spintronics, the first and  third terms are features of the Rashba model and are responsible for the
 EE and the SHE, respectively. To appreciate this,  we recall that under very general conditions, the coupling between charge and spin currents can be described~\cite{Dyakonov07} in terms of a single parameter $\gamma$ -- the spin Hall angle --  which must be evaluated from a microscopic model. 
If  we now consider a definite geometry where 
the external electric  field (which is present in both effects) is applied along the $x$ direction, Eqs.~(\ref{SpinContinuity}-\ref{SU2ChargeCurrent}) read
\begin{eqnarray}
\dot {\delta S^{y}} &=&-2m\alpha J_{y}^{z}-\frac{\delta S^{y}}{\tau_{EY}}+\chi_s\dot B^y
\label{First},\\
J^z_y&=&2m\alpha D \delta S^y +\sigma_s E^z_y+ \gamma  J_x\label{Second},\\
J_x& = & \sigma_D E_x -\gamma J^z_y\,.\label{Third}
\end{eqnarray}
In the first equation $\chi_s\dot B^y$ is the spin injection rate.  For the pure Rashba model the spin Hall angle is $\gamma =-2m\alpha^2 \tau$~\cite{Schwab10}, 
which  corresponds to  the $\omega_c\tau$ of the classical magneto-transport theory, where the cyclotron frequency, $\omega_c$,  is replaced by $2 m\alpha^2$.    When $\lambda\neq 0$, the parameter $\gamma$ gets additional contributions proportional to $\lambda$, due to the so-called side-jump and skew-scattering mechanisms. We refer to Ref.~[\onlinecite{Raimondi12}] for details. Clearly the spin Hall effect is a consequence of the {\it Hall}-like term with $\sigma_{SHE}=\gamma \sigma_D$. 

Solving the  coupled equations~(\ref{First})-(\ref{Third})  yields expression for $\delta S^y$, $J^z_y$ and $J_x$, which capture the phenomenology of  the direct and inverse Edelstein effects and spin Hall effects, including the effects of extrinsic impurity scattering,  which are quite non-intuitive in the case of the DEE (see Refs.~[\onlinecite{Raimondi12, Gorini12}]).  
In particular, setting $E_x=E^z_y=0$ yields
\begin{eqnarray}
\delta S^{y} & = & \frac{\tau_s\chi_s  \dot B_{y}}{1-i\omega\tau_s}\label{ssy}\\
J_{y}^{z} & = & 2 m\alpha D\frac{\tau_s\chi_s \dot B_{y}}{1-i\omega\tau_s}\\
J_{x} & = & -\frac{2\pi}{e} \alpha\sigma_{SHE}\frac{\tau_s\chi_s \dot B_{y}}{1-i\omega\tau_s}
\end{eqnarray}
where the total relaxation time is given by $\tau_s=1/(\tau_{EY}^{-1}+\tau_{DP}^{-1})$ with $\tau_{DP}^{-1}=(2m\alpha)^2D$ the standard D'yakonov-Perel' spin relaxation time. In the low frequency limit, the coefficient of the IEE reads
\be
\sigma_{IEE}=\frac{2}{ e}\alpha m\tau_s\sigma_{SHE}.
\ee
%

In the most interesting regime in which the Rashba spin precession dominates extrinsic processes, the EY spin relaxation process is negligible and the spin Hall conductivity is given by $\sigma_{SHE}\simeq -(e/8\pi)(4\tau/\tau_{DP})$. 
In this regime, which is directly relevant to the experiments of Rojas S\'anchez {\it et al.}~\cite{Sanchez13},  we obtain 
\be
\sigma_{IEE}=-\alpha m\tau/\pi.
\ee
%
This is the result that would have been obtained by computing the anomalous part of the current $J_x$ in the presence of Rashba coupling, using for the expectation value of  $\delta S^y$ the non-equilibrium spin polarization injected by the source $\chi_s\dot B^y$.  Although derived for the diffusive regime, this result remains valid in the ballistic regime, due to the cancellation of the spin relaxation rate contained in $\delta S_y$ against the one contained in the denominator of  $\sigma_{SHE}$.

{\it Discussion of experiments} --
In a recent experiment, Rojas S\'anchez {\it et al.}~\cite{Sanchez13} have observed the inverse Edelstein Effect at the Ag/Bi interface.  The Ag/Bi interface hosts a 2DEG of surface density $n \simeq6\times10^{13}$ cm$^{-2}$,  corresponding to a  Fermi wave vector $k_F \simeq 0.2$ \AA$^{-1}$~\cite{Ast07,Bian12}.  These electrons resides in states bound to the interface and propagate only in the plane of the interface with an effective mass   $m^* \simeq 0.35 m$~\cite{Ast07}.   They are subjected to an unusually large Rashba spin-orbit field, ${\hbar}{\alpha} \simeq 1$~eV \AA, and   they are well described by the Rashba 2DEG Hamiltonian of Eq.~(\ref{RH}).    In practice, rather than using a time-dependent magnetic field as we proposed above, Rojas S\'anchez {\it et al.} inject the non-equilibrium spin polarization by a spin current generated by ferromagnetic resonance of a remote NiFe layer. Since the injected spin current flows perpendicular to the interface it does not propagate but leads to a non-equilibrium  spin accumulation at the interface.  Thus,   the observed in-plane charge current cannot be explained by the ISHE of  the interfacial electron gas (the signal is demonstrated to be not due to the ISHE in the bulk Ag or Bi~\cite{Sanchez13}).  We now apply our theory to the analysis of this experiment. 

Obviously, in the absence of external magnetic field, the equilibrium distribution is unpolarized, i.e., $\delta S^y\equiv S^y$. Moreover, the spin pumping term in Eq.\,(\ref{First}), $\chi_s\dot B_y$, should be replaced by the injected spin current density $J_s^y$ (polarized along $y$ direction). 
Notice that this latter spin current density is three-dimensional (i.e., related to number of electrons per unit volume) in contrast to the charge current density, which is a surface density. Hence the ratio $J_x/J_s^y$ must have the dimensions of a length.
Therefore, the induced charge current is expressed by
\be
J_x =  -\frac{2\pi}{ e}\alpha\tau_s\sigma_{SHE} J_s^y\xrightarrow[\text {dominant}]{\text {Intrinsic}}\alpha \tau J_s^y. \label{jxjsy}
\ee
%
The result in the intrinsic limit is similar to that suggested by the simple two-band model in the experimental paper, $J_x=\alpha \tau_s J_s^y$~\cite{Sanchez13}. 
 Whereas  in Ref. \cite{Sanchez13} is suggested  that the relaxation time in this formula  effectively takes into account the coupled spin-momentum dynamics, our theory provides a full microscopic derivation of it. In particular, our
theory shows that the relaxation time present in the ratio between induced charge current and injected spin current should be the momentum relaxation time, 
even though the magnitude of the spin polarization itself is proportional to the spin relaxation time [see Eq.\,(\ref{ssy})]. This property is also demonstrated by our calculation from Kubo's formula (not shown). The underlying physics is that the generation of charge current from a spin polarization is mediated by an in-plane spin current (generated by precession in the Rashba field -- see Fig. 1) which is therefore proportional to the SHE coefficient, introducing a factor $\tau/\tau_{DP}$ as shown in Eq.\,(\ref{jxjsy}). 
With the measured value of $\lambda_{IEE}=\alpha \tau=0.3$ nm, we estimate $\tau= 2 \times 10^{-15}$ s and $\tau_s =3 \times 10^{-15}$ s, which puts us at the borderline of the spin-diffusive regime\footnote{We should emphasize that, the  correct expression to extract the IEE coefficient, which satisfies the Onsager relation discussed above, should be $J_x/(J_s^y\chi_s)$.}.

{\it Conclusions}--
In this paper we have provided a formal definition of the IEE in terms of the standard Kubo response functions. 
For the case of a 2DEG with Rashba spin-orbit interaction we have shown how the IEE arises as a combination of the $z$-spin current
flowing along the $y$-direction due to an non-equilibrium $S^y$ polarization and of the ISHE mechanism which yields, in turn,  a $x$-flowing charge current. Explicit results have been shown in the diffusive regime where we have used the theoretical framework of the $SU(2)$ formulation
for linear-in-momentum spin-orbit coupling. Finally, we have compared our theory with recent experimental results.

We thank Albert Fert and Cosimo Gorini for stimulating discussions.
We  acknowledge support  from NSF Grant No. DMR-1104788 (KS) and from the SFI Grant 08-IN.1-I1869 and the Istituto Italiano di Tecnologia under the SEED project
grant No. 259 SIMBEDD (GV).   RR acknowledges partial support from EU through Grant. No. PITN-GA-2009-234970.


\begin{thebibliography}{100}
\bibitem{footnote1} There is nevertheless a formal difference between the two processes.  In the current-injection case, the instantaneous equilibrium current vanishes because of the exact cancellation between the paramagnetic and the diamagnetic components of the equilibrium current.   In the spin-injection case the instantaneous equilibrium spin density is purely paramagnetic and does not vanish.
\bibitem{footnote_EY} A microscopic derivation of this relaxation time approximation is possible~\cite{Raimondi10}.
\bibitem{footnote_ddh}{In this paper  $J^a_i $ is the conventional spin current.  We note that this differs from the alternative definition of the spin-current as time derivative of the spin dipole, which was introduced a few years ago by Shi et al.~\cite{Shi06} in their discussion of Onsager's relations.   The present definition is perfectly suitable, provided the derivatives that appear in the continuity equation and the diffusion equation for the spin are replaced by the corresponding SU(2) covariant derivatives~\cite{Raimondi12, Gorini12}.}
\end{thebibliography}
\end{document}